\documentclass[twocolumn,journal]{IEEEtran}
\usepackage[T1]{fontenc}
\usepackage{amsmath}
\usepackage{graphicx}
\usepackage[unicode=true,
 bookmarks=true,bookmarksnumbered=true,bookmarksopen=true,bookmarksopenlevel=1,
 breaklinks=false,pdfborder={0 0 0},backref=false,colorlinks=false]
 {hyperref}
\hypersetup{pdftitle={Your Title},
 pdfauthor={Your Name},
 pdfpagelayout=OneColumn, pdfnewwindow=true, pdfstartview=XYZ, plainpages=false}
\usepackage{breakurl}

\makeatletter

\providecommand{\tabularnewline}{\\}

\usepackage[caption=false,font=footnotesize]{subfig}
\usepackage{amsmath}

\@ifundefined{showcaptionsetup}{}{%
 \PassOptionsToPackage{caption=false}{subfig}}
\usepackage{subfig}
\makeatother

\begin{document}

\title{Physical Layer Key Generation for Secure Power Line Communications}

\author{Federico~Passerini,~\IEEEmembership{Student Member,~IEEE,} and~Andrea~M.~Tonello,~\IEEEmembership{Senior~Member,~IEEE}\thanks{Federico Passerini and Andrea M. Tonello are with the Embedded Communication
Systems Group, University of Klagenfurt, Klagenfurt, Austria, e-mail:
\{federico.passerini, andrea.tonello\}@aau.at.}}
\maketitle
\begin{abstract}
Leakage of information in power line communication networks is a threat
to privacy and security both in smart grids and in-home applications.
A way to enhance security is to encode the transmitted information
with a secret key. Relying on the channel properties, it is possible
to generate a common key at the two communication ends without transmitting
it through the broadcast channel. Since the key is generated locally,
it is intrinsically secure from a possible eavesdropper. Most of the
existing physical layer key generation techniques have been developed
for symmetric channels. However, the power line channel is in general
not symmetric, but just reciprocal. Therefore, in this paper, we propose
two novel methods that exploit the reciprocity of the power line channel
to generate common information at the two intended users. This information
is processed through different quantization techniques to generate
secret keys. To assess the security of the generated keys, we analyze
the spatial correlation of the power line channels and verify the
low correlation of the possible eavesdropping channels. The two proposed
methods are tested on a measurement dataset. The results show that
the information leaked to possible eavesdroppers has very low correlation
to any secret key.\end{abstract}


\section{Introduction}

\IEEEPARstart{I}{nformation} in networks where the communication
mean is shared is always at risk, since both authorized and illegitimate
users are given physical access to the network. Malicious users have
therefore a chance to jeopardize the privacy of other users or, conversely,
to send false information throughout the network. Typical examples
of networks where such risk is particularly threatening are wireless
networks and power line networks (PLNs). 

In such physical broadcast (PB) networks, conversely from classical
computer networks, a malicious user can perform attacks on all the
stacks of the ISO/OSI model, including the MAC and physical layer
\cite{7467419}. In particular, the physical layer (PHY) comes to
play an important role in both planning attacks to the network and
defensive strategies. In fact, since the physical medium is shared,
every input into the network has an effect on the system outputs.
If the network system can be modeled, then its properties can be used
with both malicious or aiding intent.

The wireless community has extensively relied on the properties of
the physical channel to pursue research and identify methods for information
security. From an information theoretic point of view, it is possible
to guarantee secure transmission when the intended communication channel
has higher capacity than the eavesdropper one, by transmitting information
at a sufficiently high rate \cite{5751298}. However, some eavesdropper
channels might have higher capacity than the intended one. For this
reason, different techniques to enhance security have been conceived
in the communication theory area, which include  secret key generation,
prefiltering and coding techniques \cite{7448884,5751298,Bloch:2011:PSI:2829193}.
These techniques rely on different properties of the wireless communication
channels to restrict the information leakage to any possible unauthorized
receiver. Such properties include the channel randomness both in time
and in frequency domain and, especially in time-division duplexing
systems, its symmetry. In fact, if the channel between two users is
symmetric, the randomness of the channel is common to the two users,
i.e. they have access to the same information. This property is particularly
useful for the secret key generation techniques. The key generation
process includes the common information, which is unknown to an eventual
eavesdropper, thus drastically enhancing the security of the produced
key. 

On the other side, in the context of power line transmission and distribution
networks, attacks and defensive strategies are normally based on system
theory. In this case, the network is modeled as a dynamic system that
describes the power flow. Attacks of different kind aim at altering
the perception of the state of the network, which in turn might bring
to a network failure \cite{6016202}. In any case, informative signals
need to circulate through the network, therefore a resilient communication
architecture would enhance the PLN security. However, to our knowledge
there is very limited literature about physical-layer secure communications
in PLNs, and it focuses only on information theoretic analysis \cite{6525863,6827060,0f4de0c7518b112301518c0d527500de,6812346}. 

In this regard, Power Line Communications (PLC) is a well established
communication technology in PLNs \cite{7467440}. This technology
already provides a form of security in the fact that it uses a communication
mean, the power line cables, that is owned by the utility and therefore
not accessible by everybody. However, an unauthorized user might be
able to get physical access to the network, or the utility might not
want to share some information with part of the network users. Therefore,
additional security measures have to be provided. Since the PLC physical
channel has some properties in common with the wireless channel, it
makes sense to explore physical-layer security (PLS) techniques developed
for wireless in the case of PLC. However, it has been shown that the
PLC channel, conversely from the wireless one, is rather deterministic,
in general not symmetric \cite{880817,7476282}, and moreover has
different statistical properties \cite{0f4de0c7518b112301518c0d527500de}.

In this paper, we propose a thorough analysis of the properties of
the PLC channel in order to investigate under what conditions PLS
techniques, which exploit common information at the two legitimate
users, developed for wireless apply also to PLC. However, since the
PLC channel is in general non symmetric, most of the known PLS algorithms
cannot be applied to it. In fact, to the authors knowledge, very limited
work exists on PLS in non-symmetric channels \cite{7523206}. In order
to overcome this limit, we make use of the fact that the PLC channel
is reciprocal to investigate what channel state information (CSI)
is known to two legitimate users independently from each other at
any given time. In this context, we propose two new methods to retrieve
common CSI. The first method involves the analysis of the multipath
signal propagation in reciprocal channels. The second method relies
on the exchange of a minimal amount of information between the two
ends, which is however insufficient to a possible eavesdropper for
decrypting the key. The CSI obtained with the proposed methods can
be consequently used to generate cryptographic keys separately at
the two communication ends. To this purpose, we process the CSI with
various quantization techniques and show the reliability of the generated
keys. We also analyze the spatial correlation in PLNs, in order to
verify the level of security of the obtained CSI against possible
eavesdroppers. Although our investigation focuses on PLN, the proposed
CSI retrieval methods are common to every reciprocal network, including
any kind of passive wired and wireless network.

The rest of the paper is organized as follows. In Section \ref{sec:Channel-based-security-approache},
we briefly summarize the existing PLS techniques based on the properties
of symmetric channels. In Section \ref{sec:Properties-of-the}, we
analyze in what cases the PLC channel can be considered symmetric.
The two algorithms for PLS in reciprocal channels are proposed in
Section \ref{sec:Enhancing-security}, while extended results are
presented in Section \ref{sec:Results}. Conclusions follow in Section
\ref{sec:Conclusions}.

\section{Channel-based security approaches in Wireless\label{sec:Channel-based-security-approache}}

\begin{figure}[tb]
\begin{centering}
\includegraphics[width=1\columnwidth]{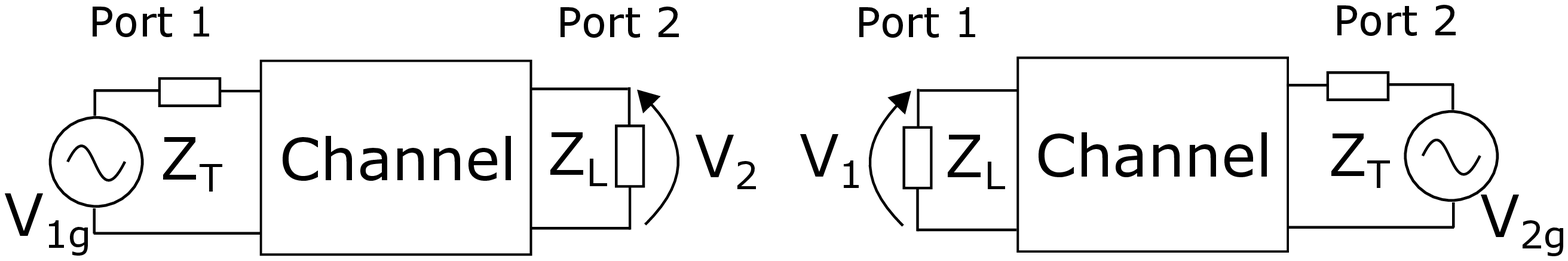}
\par\end{centering}

\caption{Traditional communication scheme, with generator either on Port 1
or on Port 2.\label{fig:Traditional-communication-scheme}}
\end{figure}
 Most of the PHY key generation techniques in wireless communications,
especially the channel-based ones, rely on the symmetry of the channel
\cite{Bloch:2011:PSI:2829193}, i.e. the fact that for every set of
currents and voltages at the two communication ends that satisfies
a certain system of relations, the set obtained by exchanging the
transmitter and the receiver satisfies the same relations. In short,
this means that the CSI is identical for both links. 

In this paper, we model the transmitter with its Thevenin or Norton
equivalent with \textit{transmit} impedance $Z_{T}$, and the receiver
with its \textit{receive} impedance $Z_{L}$. Moreover, we consider
the physical channel to be a system accessible in two ports, Port
1 and Port 2, where the transmitter and the receiver are attached
(see Fig. \eqref{fig:Traditional-communication-scheme}). The channel
transfer functions (CTFs) $H_{1}$ and $H_{2}$ defined as\footnote{We remark that \eqref{eq:H} and the rest of the equations presented
in this paper are function of the frequency. This dependency is omitted
in the notation for simplicity.} 
\begin{equation}
H_{1}=\frac{V_{2}}{V_{1g}}\quad H_{2}=\frac{V_{1}}{V_{2g}}\label{eq:H}
\end{equation}
(see Fig. \ref{fig:Traditional-communication-scheme}) are equal in
symmetric channels. Therefore, when the receiver estimates for example
$H_{1}$ of the forward link, it directly knows also $H_{2}$ of the
reverse link, without need of further communication\footnote{The wireless literature often refers to this property as due to the
reciprocity of the channel. This is technically incorrect, because
in reciprocal networks the CTF is not forcefully the same in the two
directions. Although the wireless channel is indeed reciprocal, it
is also in most of the cases symmetric, as we will explain in Section
\ref{sec:Enhancing-security}.}. Such property serves as source of common randomness from which the
parties can generate secret keys. An eavesdropper normally experiences
a physical channel that is independent of that of the legitimate users.
Therefore, the generated keys are intrinsically secure. In the following,
we name the two legitimate parties Alice (A) and Bob (B) respectively,
and we name the eavesdropper Eve (E). We also assume that Eve is a
passive attacker, i.e. she just overhears the channel. 

The basic idea of channel-based key generation approaches is for Alice
and Bob to obtain very correlated observations of the channel via
channel training, then to apply key generation methods that rely on
the correlated observations and public discussion \cite{7393435}.
 From an information-theoretic perspective, the key generation procedure
can be described as follows:
\begin{enumerate}
\item \textit{Channel sensing}: Alice, Bob and Eve get the observations
of length $n$ of the CSI $X^{n}=[X_{1},\cdots,X_{n}]$, $Y^{n}=[Y_{1},\cdots,Y_{n}]$,
and $Z^{n}=[Z_{1},\cdots,Z_{n}]$ respectively, where the observations
can be performed in time, frequency, space domain or a combination
of them. 
\item \textit{Key reconciliation via public discussion}: in order to agree
on a secret key, Alice and Bob can communicate through the PB channel
and send to each other a deterministic communication sequence as follows.
They generate the random variables $U_{A}$ and $U_{B}$ respectively,
for initialization. Then, they alternatively send to each other the
two sequences $S_{A}^{k}=[S_{A_{1}},\cdots,S_{A_{k}}]$ and $S_{B}^{k}=[S_{B_{1}},\cdots,S_{B_{k}}]$,
respectively, where for each step $i$ we have $S_{A_{i}}=f_{A_{i}}\left(U_{A},X^{n},S_{B_{i-1}}\right)$
and $S_{B_{i}}=f_{B_{i}}\left(U_{B},Y^{n},S_{A_{i-1}}\right)$. At
the end of the communication step, Alice and Bob determine the respective
keys as $K_{A}=f_{A_{k+1}}(U_{A},X^{n},S_{B}^{k})$ and $K_{B}=f_{B_{k+1}}(U_{B},Y^{n},S_{A}^{k})$.
Different protocols have been proposed to implement both the reconciliation
procedure, implemented either with cascade or error correcting codes,
and the privacy amplification. An extended series of references about
this can be found in \cite{7393435}.
\end{enumerate}
By definition \cite{256484}, a secret key rate $R_{K}$ is achievable
if for every $\varepsilon>0$ and sufficiently large $n$, there exists
a public communication strategy such that\begin{subequations}
\begin{align} 
Pr\left\{K_A\neq K_B\right\} & < \varepsilon \label{eq:a}\\ 
\frac{1}{n} I\left(K_A;S_A^k,S_B^k,Z^n\right) & < \varepsilon \label{eq:b}\\
\frac{1}{n} H\left(K_A\right) & > R_K-\varepsilon \label{eq:c}\\ 
\frac{1}{n}\log\left|\mathcal{K}\right| & < \frac{1}{n} H\left(K_A\right)+\varepsilon, \label{eq:dudud}
\end{align} 
\label{eq:def}%
\end{subequations}where $H(\cdot)$ and $I(\cdot)$ denote the entropy and mutual information
operators and $\mathcal{K}$ is the key alphabet. Equation \eqref{eq:a}
means that $K_{A}$ and $K_{B}$ are \textit{equal}, \eqref{eq:b}
ensures that no information is leaked to Eve and \eqref{eq:dudud}
indicates that the generated key is uniformely distributed. It is
clear from \eqref{eq:def} that the possibility of generating at least
one ($R_{K}>1$) or multiple keys is based on three characteristics
of the PB medium: the\textit{ temporal variation }(i.e. the randomness)
and the \textit{correlation of the CSI} between Alice and Bob, and
the \textit{spatial decorrelation} of Eve. These three characteristics
are fulfilled in many wireless scenarios, where the channel varies
frequently, it is symmetric and the users typically experience uncorrelated
multipath fading. This practically means, respectively, that $n$
(considering observations in time) is low, $X^{n}$ and $Y^{n}$ are
very correlated, which guarantees a fast convergence for the condition
\eqref{eq:a}, and they are both uncorrelated with $Z^{n}$, which
guarantees the convergence of \eqref{eq:b}.

In the following, we analyze how the characteristics of the power
line medium can be used to retrieve highly correlated CSI among Alice
and Bob. Moreover, we discuss the physical constraints that limit
the achievable $R_{K}$ in PLN. Considering the system model introduced
in Fig. \ref{fig:Traditional-communication-scheme}, we assume Alice
to be branched at Port 1 and Bob or Eve to be branched at Port 2,
depending on which CTF is of interest.

\section{Symmetries of the Power Line Channel\label{sec:Properties-of-the}}

\begin{figure}[tb]
\subfloat[\label{fig:a}]{\begin{centering}
\includegraphics[width=1\columnwidth]{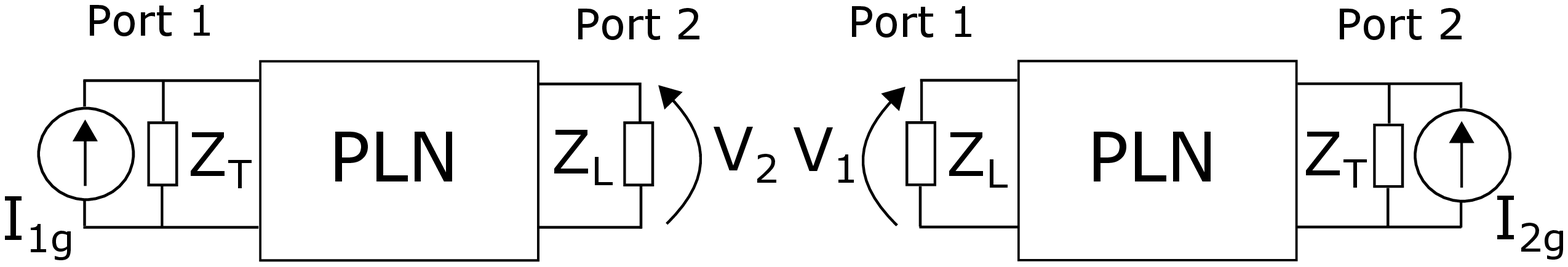}
\par\end{centering}

}

\subfloat[\label{fig:b}]{\begin{centering}
\includegraphics[width=1\columnwidth]{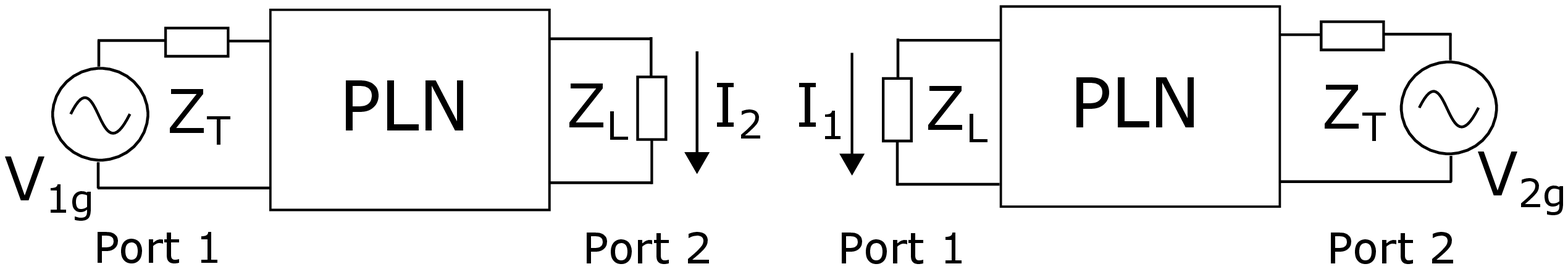}
\par\end{centering}

}\caption{Trans-resistance a) and trans-conductance b) communication schemes
in PLN, with generator either on Port 1 or on Port 2.\label{fig:Current-a)-and}}

\end{figure}

In this section, we present under which conditions the power line
channel is symmetric. Under these conditions, the existing PLS techniques
developed for wireless can be similarly applied to PLNs.

It has been shown in \cite{880817} that the power line channel is
symmetric if the impedance $Z_{T}$ at the transmission side is equal
to the load impedance $Z_{L}$ (see Fig. \ref{fig:Traditional-communication-scheme}).
Similarly, this condition applies to the wireless channel and to any
other kind of passive network. However, while in wireless systems
both $Z_{T}$ and $Z_{L}$ are set to the same value (usually 50$\Omega$)
to maximize the power transmitted and received, the situation is different
in PLC. 

In classical half-duplex PLC systems, maximum communication rate is
obtained by maximizing the transferred voltage, or more in general
the SNR at the receiver \cite{7407357}. Therefore, PLMs are usually
equipped with $Z_{T}\sim1\Omega$, $Z_{L}\sim10$k$\Omega$ and a
switch that selects the correct impedance based on the link status
\cite{NB_frontend}. This renders the channel highly non symmetric.
On the other hand, in the recently proposed in-band full duplex PLC
technology, some front-end transceiver architectures use the same
equivalent impedance both for transmission and reception chains \cite{isplc2018}.
Therefore, if the two communication ends are equipped with modems
that use an equivalent impedance with the same value, the channel
is symmetric. 

A third communication architecture, which has been not yet proposed
in the context of PLC, can be considered. It relies on the fact that
the PLC channel is reciprocal \cite{7476282} to get symmetric CSI.
In fact, in any reciprocal two-port network the following holds true
\cite{Pozar}: 
\begin{itemize}
\item When the current $I_{g}$ is applied to any of the two ports, the
open circuit voltage measured at the other port is the same. Referring
to Fig. \ref{fig:a}, this means that the ratio $Z_{21}=V_{2}/I_{1g}$
is equal to the ratio $Z_{12}=V_{1}/I_{2g}$ obtained when the two
ports are inverted, under the condition $Z_{T}=Z_{L}=\infty$
\item When the voltage $V_{g}$ is applied to any of the two ports, the
short circuit current measured at the other port is the same. Referring
to Fig. \ref{fig:b}, this means that the ratio $Y_{21}=I_{2}/V_{1g}$
is equal to the ratio $Y_{12}=I_{1}/V_{2g}$ obtained when the two
ports are inverted, under the condition that $Z_{T}=Z_{L}=0$. 
\end{itemize}
Therefore, it is possible to obtain symmetric transmission of signals
considering the trans-impedance $Z$ or the trans-admittance $Y$
of the network instead of the classical voltage transfer function
(see Fig. \ref{fig:Current-a)-and}). However, the values of the transmit
and receive impedances under which this property strictly holds are
ideal and far from the common values of $Z_{T}$ and $Z_{L}$. 

\begin{figure}[tb]
\begin{centering}
\includegraphics[width=0.9\columnwidth]{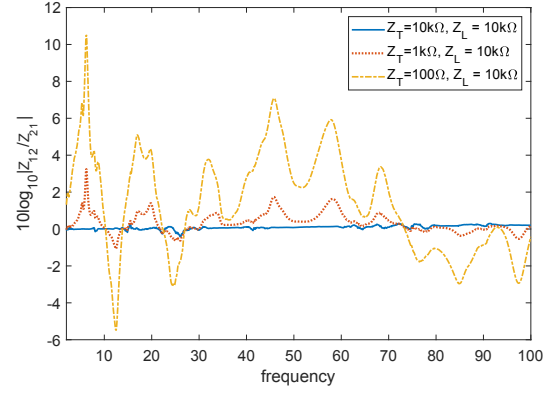}
\par\end{centering}

\caption{Symmetry of the trans-impedance $Z$ for different values of $Z_{T}$
and $Z_{L}$.\label{fig:Symmetry-of-}}

\end{figure}
 We now consider less ideal conditions, taking the trans-impedance
case as an example. We name $Z_{12}$ the one obtained from transmission
from Port 2 to Port 1 and $Z_{21}$ the opposite one. We also fix
$Z_{L}$ to $10$k$\Omega$, as usual in PLM receivers and modify
the value of $Z_{T}$. Fig. \ref{fig:Symmetry-of-}, which is obtained
from a dataset as discussed in Section \ref{sec:Results}, shows that
for low values of $Z_{T}$ the trans-impedance is highly asymmetric.
The symmetry increases with the value of $Z_{T}$, and when $Z_{T}$
reaches $10$k$\Omega$, the trans-impedance is essentially symmetric.
This condition would be practically implementable in power line modems,
by driving the line with a current instead of a voltage \cite{1051875}
and using a classical voltage receiver. Even though not shown, a similar
result is obtained in the trans-admittance case when $Z_{T}$ and
$Z_{L}$ are close to or less then 1$\Omega$. Implementing this solution
in power line modems would imply to send a voltage signal using a
classical transmitter and to receive a current signal over a very
small impedance. 

When $Z_{T}$ and $Z_{L}$ have values that are far from ideality,
the following method can be used to obtain symmetric CSI at the two
communication ends. Referring to the trans-impedance case, we point
out that if a circuit is adopted to measure the PLN input impedance
$Z_{in_{k}}$ at the Port $k$ that is defined as
\begin{equation}
Z_{in}=\frac{V_{k}}{I_{k}},\label{eq:zin}
\end{equation}
then, relying the voltage and current divider equations, we can write
\begin{equation}
I_{1i}=\frac{Z_{T}}{Z_{in_{1}}+Z_{T}}I_{1g}\quad V_{2oc}=\frac{Z_{in_{2}}+Z_{L}}{Z_{L}}V_{2},\label{eq:3}
\end{equation}
\begin{equation}
I_{2i}=\frac{Z_{T}}{Z_{in_{2}}+Z_{T}}I_{2g}\quad V_{1oc}=\frac{Z_{in_{1}}+Z_{L}}{Z_{L}}V_{1},\label{eq:4}
\end{equation}
for the transmission from Port 1 and Port 2 respectively. $I_{i}$
is the actual current entering the network and $V_{oc}$ is the open
circuit voltage at the receiver. These two new quantities are the
equivalent of the injected current and received voltage if the transmit
and receive impedance were infinte. Similarly, in the trans-admittance
case we have
\begin{equation}
V_{1i}=\frac{Z_{in_{1}}}{Z_{in_{1}}+Z_{T}}V_{1g}\quad I_{2cc}=\frac{Z_{in_{2}}+Z_{L}}{Z_{in_{2}}}I_{2},\label{eq:5}
\end{equation}
\begin{equation}
V_{2i}=\frac{Z_{in_{2}}}{Z_{in_{2}}+Z_{T}}V_{2g}\quad I_{1cc}=\frac{Z_{in_{1}}+Z_{L}}{Z_{in_{1}}}I_{1},\label{eq:6}
\end{equation}
where $V_{i}$ and $I_{cc}$ are the equivalent of the injected current
and received voltage if the transmit and receive impedance were zero.
This means that \eqref{eq:3},\eqref{eq:4},\eqref{eq:5} and \eqref{eq:6}
allow us to reproduce the conditions for symmetry in the respective
systems. In fact, the resulting trans-impedances $Z_{21}^{'}=V_{2oc}/I_{1i}$
and $Z_{12}^{'}=V_{1oc}/I_{2i}$, as well as the trans-admittances
$Y_{21}^{'}=I_{2cc}/V_{1i}$ and $Y_{12}^{'}=I_{1cc}/V_{2i}$, are
respectively equal, independently of the actual values of $Z_{T}$
and $Z_{L}$ used. 

In conclusion, a symmetry can be derived as explained also using classical
values of output and load impedances in PLMs. A possible drawback
of this method is that the receiver needs to know both $Z_{in_{1}}$
and $Z_{in_{2}}$. Hence, the value of $Z_{in_{1}}$ or $Z_{in_{2}}$
needs to be transmitted through the public channel with risk of eavesdropping.
However, a possible eavesdropper would not have access to the values
of $V_{1}$ or $V_{2}$ in the trans-impedance case or to $I_{1}$
or $I_{2}$ in the trans-admittance case, which are a trait of the
intended receiver. Therefore, sharing information about the channel
input impedance at the transmitter and at the receiver, does not directly
enable an eventual eavesdropper to estimate for example $Z_{21}^{'}$.
This approach is further elaborated and discussed in Section \ref{sub:Transmission-matrix-technique},
where we do not limit to trans-impedance or trans-admittance architectures,
but we generalize this method to any kind of communication architecture.

\section{Key generation in half-duplex PLC\label{sec:Enhancing-security}}

In this Section, we propose two techniques to get common information
at the transmitter and the receiver with minimal exchange of data.
Both techniques rely on the fact that the PLC channel is reciprocal,
as discussed before.

\subsection{Time-domain symmetry technique (TDST)\label{sub:Time-domain-symmetry-technique}}

\begin{figure}[tb]
\begin{centering}
\subfloat[\label{fig:a-1}]{\centering{}\includegraphics[width=1\columnwidth]{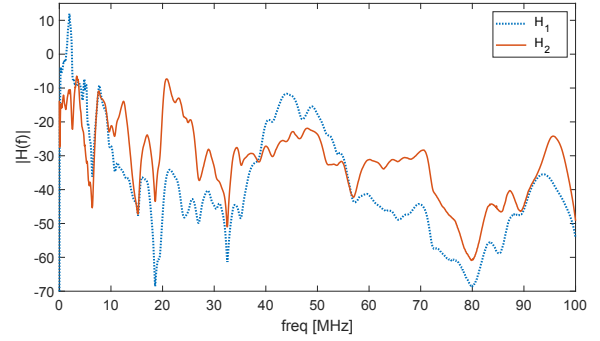}}
\par\end{centering}

\subfloat[\label{fig:b-1}]{\begin{centering}
\includegraphics[width=1\columnwidth]{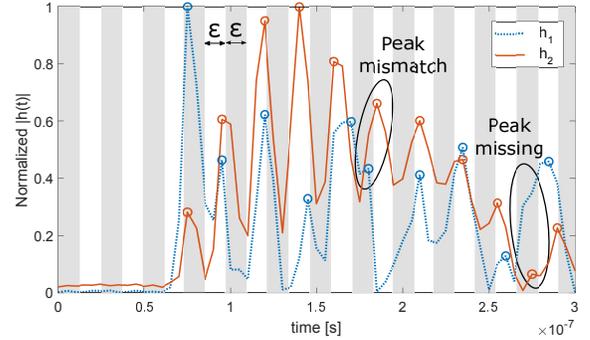}
\par\end{centering}

}\caption{Example of a PLC channel transfer function in the two directions,
in frequency a) and time b) domain.\label{fig:Example-of-a}}
\end{figure}
Considering a generic two port network, which in our case represents
the PLN, the transmission matrix is defined as \cite{Pozar}
\begin{equation}
\left[\begin{array}{c}
V_{1}\\
I_{1}
\end{array}\right]=\left[\begin{array}{cc}
A & B\\
C & D
\end{array}\right]\left[\begin{array}{c}
V_{2}\\
I_{2}
\end{array}\right],\label{eq:ABCD}
\end{equation}
where the subscripts 1 and 2 stand for the relative port. When the
system is reciprocal, which is always the case in passive networks,
the following relation holds true
\begin{equation}
AD-CB=1.\label{eq:condition}
\end{equation}
With this condition, the transmission matrix in the opposite direction
becomes 
\begin{equation}
\left[\begin{array}{c}
V_{2}\\
I_{2}
\end{array}\right]=\left[\begin{array}{cc}
D & B\\
C & A
\end{array}\right]\left[\begin{array}{c}
V_{1}\\
I_{1}
\end{array}\right].\label{eq:DBCA}
\end{equation}
As shown in Appendix \ref{sec:Wide-sense-symmetry-of}, the time-domain
response of \eqref{eq:ABCD} and \eqref{eq:DBCA} is not strictly
symmetric but wide-sense symmetric. This means that the multipath
response of the channel is characterized by peaks that are in the
same position both when the signal travels from Port 1 to Port 2 and
vice versa. However, the amplitude of the peaks and their shape are
in general different, thus the PLC channel is not strictly symmetric.
As an example, Fig. \ref{fig:Example-of-a} shows the frequency and
time domain response of a typical PLC channel in the two communication
directions. The frequency domain response is far from symmetric, even
though a certain degree of correlation still exists. The wide-sense
symmetry in time domain appears clearly in Fig. \ref{fig:b-1}. Even
though the amplitude of the peaks in the two cases is rather different,
we see that their position is the same. The mismatches are due mainly
to two reasons. On one hand, high peaks might render lower peaks that
are close to them undetectable. On the other hand, the peak detection
algorithm and the bandwidth of the signal deeply influence the estimation
of the peak presence and position. 

One way to compensate these errors and to construct a key is to divide
the time domain response $h$ (or part of it) in $N$ blocks, each
with duration $\varepsilon$ (white and gray stripes in Fig. \ref{fig:b-1}).
A binary key with $N$ elements is generated at each node, with all
values initially set to 0. After channel estimation and peak detection,
every key element is set to one if at least one peak is detected within
its time block, so that the binary key $K$ is generated. This method
can be further refined by limiting the peak search to the first $M$
blocks set to one. The limit is set because, due to the multipath
and the smoothing effect of the channel, the density of the peaks
tends to increase and their granularity tends to decrease with the
time index, respectively. This means that every possible $K$ would
have a lot of ones towards the end of the sequence, which results
in high similarity between different keys. Converserly, when the limit
to the first $M$ ones is applied, there are higher chances that the
position of the ones in keys generated from Alice and Eve are in different
positions. Finally, key reconciliation procedures, such as Slepian-Wolf
coding \cite{Bloch:2011:PSI:2829193} can be run as presented in Section
\ref{sec:Channel-based-security-approache} to agree on the final
key.

A drawback of the TDST is the generation rate of new keys, which is
very low or even zero. This is because the position of the peaks in
the time domain response is due to the topological structure of the
network. Thus, the key would change only when a topology variation
occurs. Small physical variations of the channel, due for example
to its periodic time variant nature \cite{1650333} or to impedance
changes at the terminations, do not in general affect the presence
or the position of peaks in the time domain channel response. The
topology is only modified when a power switch is activated to route
the power flow to a different section of the grid or when an anomaly
like a fault or a strong impedance change occurs \cite{SGSI}. 

In the case of transmission and medium voltage distribution networks,
topological variations might occur from hours to weeks one from another.
In the case of indoor or low voltage distribution networks, the topology
of PLNs is fixed unless an anomaly occurs, therefore each communication
pair can generate just one code. Since frequent channel changes are
needed to prevent eventual eavesdroppers to retrieve the communication
key, this key generation technique is prone to be decrypted over a
long time period. Increased security could be obtained, for example,
by combining the TDST with classical cryptographic methods to periodically
refresh the key.

\subsection{Transmission matrix technique (TMT)\label{sub:Transmission-matrix-technique}}

Taking as starting point the normalization procedure presented in
Section \ref{sec:Properties-of-the}, we can extend it to derive the
full transmission matrix of the communication link. For this purpose,
we assume the power line modems to be enabled to provide an estimate
$\tilde{H}$ of the frequency response $H$ and $\tilde{Z_{in}}$
of the channel input impedance $Z_{in}$, respectively \cite{7994720}.

Since the parameters $A$, $B$, $C$ and $D$ of the transmission
matrix are the same in the two directions, their estimation at one
communication end would enable the complete electrical characterization
of the channel in both directions. Relying on \eqref{eq:H},\eqref{eq:zin},\eqref{eq:ABCD}
and \eqref{eq:DBCA} we can write the following equations
\begin{equation}
Z_{in1}=\frac{V_{1}}{I_{1}},=\frac{A+\frac{B}{Z_{L}}}{C+\frac{D}{Z_{L}}},\label{eq:Zin1}
\end{equation}
\begin{equation}
H_{1}=\frac{V_{2}}{V_{1g}}=\frac{Z_{L}}{Z_{L}A+B+Z_{L}Z_{T}C+Z_{T}D},\label{eq:H1}
\end{equation}
\begin{equation}
Z_{in2}=\frac{V_{2}}{I_{2}},=\frac{D+\frac{B}{Z_{L}}}{C+\frac{A}{Z_{L}}},\label{eq:Zin2}
\end{equation}
\begin{equation}
H_{2}=\frac{V_{1}}{V_{2g}}=\frac{Z_{L}}{Z_{L}D+B+Z_{L}Z_{T}C+Z_{T}A}.\label{eq:H2}
\end{equation}
The four complex unknowns $A$, $B$, $C$, $D$, can be found by
solving a system made with these four complex equations \cite{7897101}.
However, solving this system at each communication end requires information
about $Z_{in}$, $H_{1}$, $Z_{in2}$ and $H_{2}$ to be shared on
the PB channel. This would allow also any potential eavesdropper to
solve the system, resulting in no secrecy. 

On the other hand, relying on \eqref{eq:condition}, another system
of equations can be written. Considering for example the user connected
at Port 2, he can directly estimate $H_{1}$, by relying on classical
pilot signals used in communication systems \cite{1033876}, and $Z_{in2}$
with an impedance sensor. At this point, considering also \eqref{eq:condition},
only one equation is missing to derive the transmission matrix. Therefore,
the value of either $Z_{in1}$ or $H_{1}$ has to be sent through
the PB channel. If, for example, the information about $Z_{in1}$
is shared, then the user can solve the system

\begin{equation}
\begin{cases}
Z_{in1}=\frac{A+\frac{B}{Z_{L}}}{C+\frac{D}{Z_{L}}}\\
H_{1}=\frac{Z_{L}}{Z_{L}A+B+Z_{L}Z_{T}C+Z_{T}D}\\
AD-CB=1\\
Z_{in2}=\frac{D+\frac{B}{Z_{L}}}{C+\frac{A}{Z_{L}}}
\end{cases}\label{eq:system}
\end{equation}
With the estimated values of the transmission matrix, the user connected
at Port 2 can estimate $H_{2}$ using \eqref{eq:H2}. At this point,
all the PLS techniques presented in Section \ref{sec:Channel-based-security-approache}
can be applied. The same procedure applies to the user connected at
Port 1, with the transmission of information about $Z_{in1}$.

Since with this method the transmission matrix is estimated by both
legitimate users, the key can be generated from any of the transmission
matrix parameters or from a function of them. Even though some information
is shared through the PB channel, a possible eavesdropper will not
be able to correctly estimate the transmission matrix between the
legitimate users, since it will at maximum have three equations available.
When the cryptographic key is based on the degree of freedom left
to the legitimate users, then the eavesdropper has no mean to retrieve
the key.

Regarding the estimation procedure, since $H_{1}$, $Z_{in1}$ and
$Z_{in2}$ are constant as long as the transmission matrix is constant,
their best estimates $\tilde{H_{1}}$, $\tilde{Z_{in1}}$ and $\tilde{Z_{in2}}$
are given by averaging over time, assuming zero mean noise \cite{kay1993stat}.
$\tilde{A}$, $\tilde{B}$, $\tilde{C}$, $\tilde{D}$, are then simply
derived by directly solving \eqref{eq:system}. When the channel state
changes, the estimation procedure can be repeated and a new cryptographic
key is generated. 

Different methods can be proposed to quantize and arrange the selected
CSI. First of all, we consider the absolute value of the magnitude
of the selected CSI to be linearly quantized over $2^{nbits}-1$ levels
for every frequency bin. Then, we consider two ways of arranging the
data:
\begin{itemize}
\item Binary: the quantized data are converted to binary sequences with
Gray encoding to minimize the distance between symbols that are close
to each other. Each binary symbol is used as a symbol of the key.
\item Coded: the key is defined over an $2^{nbits}$-ary alphabet and each
symbol is made by the quantized value of the CSI at one frequency
bin. One symbol at the end of the key sequence accounts for the actual
value of the least significant bit. The actual key is generated by
multiplying the values of all the symbols by the last one. This method
is used to avoid data with similar shape but different amplitudes
to produce similar keys.
\end{itemize}
These two methods will be compared in Section \ref{sec:Results},
where we consider as an example the key to be derived from $\tilde{H_{2}}$.
We remark that other quantization methods are possible. However, a
thorough comparison of quantization methods is out of the scope of
this paper.

As mentioned before, the PLC channel is typically ciclostationary
with period equal to the mains semi-cicle and can be roughly subdivided
into a series of slots in which it is considered static. Such intervals
typically are in the order of some hundreds of microseconds \cite{1650333}.
Hence, the number of cryptographic keys that can be generated for
a given node pair using the proposed method is equal to the number
of time slots in the particular scenario. Since the state variations
are much higher at frequencies below 5 MHz than above, a higher number
of and less correlated keys are likely to be generated using narrow-band
PLC, which uses the spectrum 3-500 kHz, than broad-band PLC, which
uses the spectrum 2-86 MHz. Therefore, the proposed method for key
generation is expected to have the best performance when applied to
distribution networks, where PLC are used mainly within the narrow-band
spectrum.

\section{Practical results\label{sec:Results}}

The results presented in this paper are based on the measurement campaign
presented in \cite{6799210}. In this measurement campaign, the full
transmission matrix of a total of 1312 in-home channels divided in
3 sites has been measured in the frequency range from 0.1 to 80 MHz.
We chose this dataset because, to our knowledge, it is the only one
available with measurements of the full transmission matrix. However,
considering the results of other measurement campaigns conducted on
distribution grids \cite{opera}, we expect our results to be qualitatively
applicable also in the outdoor scenario.

\subsection{Channel correlation}

\begin{figure}[tb]
\begin{centering}
\includegraphics[width=1\columnwidth]{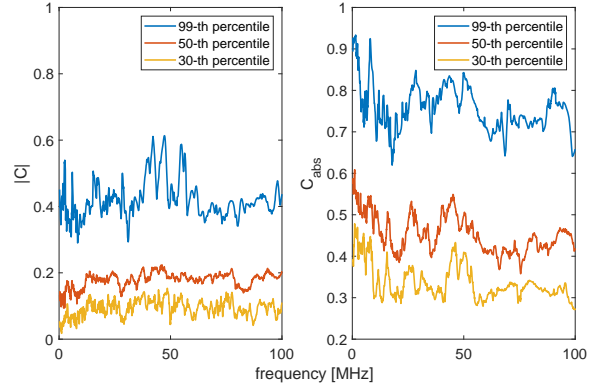}
\par\end{centering}

\caption{Correlation between the $Z_{in}$ and the CFT transmitted from the
same node. \label{fig:Correlation-between-the}}
\end{figure}
As presented in Section \ref{sec:Channel-based-security-approache},
one fundamental property to generate secure keys from the physical
channel is the strong correlation between the two forward and reverse
channels from Alice to Bob and vice-versa. At the same time, both
the Alice to Eve and Bob to Eve channels have to be as low correlated
as possible w.r.t. the two legitimate channels. Therefore, in this
section we analyze the spatial correlation of PLC channels, independently
on the key generation method used. 

A first work presented in \cite{6799210} defined the space-frequency
correlation as 
\begin{equation}
R_{\ell,m}=\frac{E_{i,j}\left[H_{i}\left(\ell\right)H_{j}^{*}\left(m\right)\right]}{\sqrt{E_{i}\left[\left|H_{i}\left(\ell\right)\right|^{2}\right]E_{j}\left[\left|H_{j}\left(m\right)\right|^{2}\right]}},\label{eq:stat_corr}
\end{equation}
where $i$ and $j$ stand for the channel realization indexes, $E\left[\cdot\right]$
is the expectation operator, $\ell$ and $m$ stand for the frequency
bin indexes, and $H$ is the CTF of a specific channel. The correlation
is computed as an expectation over all the channels that share the
same transmitter and over all the possible transmitters. The results
show that the average correlation between the channel transfer functions
from or to different outlets is rather low, but it increases to values
around 0.3 when the absolute values of $H$ are considered. 

In this work, especially in the case of the TMT, information about
the input impedance is shared, and Eve is interested in retrieving
$H_{ab}$ and $H_{ba}$ from it (see Section \ref{sub:Transmission-matrix-technique}).
Therefore, it is of interest to compute
\begin{equation}
C_{\ell,m}=\frac{E_{i,j}\left[Z_{in_{i}}\left(\ell\right)H_{Zin_{j}}^{*}\left(m\right)\right]}{\sqrt{E_{i}\left[\left|Z_{in_{i}}\left(\ell\right)\right|^{2}\right]E_{j}\left[\left|H_{Zin_{j}}\left(m\right)\right|^{2}\right]}},\label{eq:stat_corr-1}
\end{equation}
where $H_{Zin}$ and $Z_{in}$ share the same transmitter. The results
in Fig. \ref{fig:Correlation-between-the}, where we fixed $l=m$,
show that $|C|$ is on average low, but the frequency-space correlation
$C_{abs}$, computed using the absolute values, is on the other side
not negligible. Therefore, it is recommended, when generating secure
keys with the TMT, to make use of the estimated complex values and
not just their magnitude. 

\begin{figure}[tb]
\begin{centering}
\includegraphics[width=1\columnwidth]{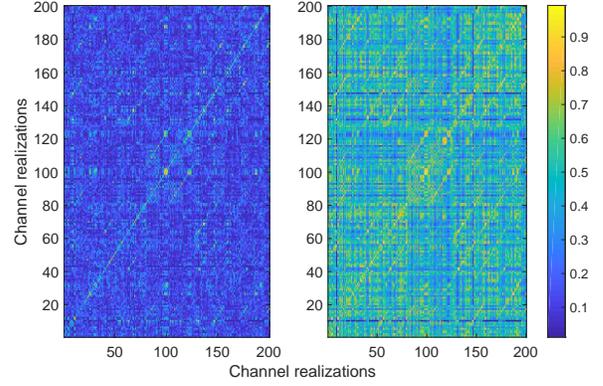}\caption{Correlation coefficients $\left|\rho^{H}\right|$ (left) and $\rho_{abs}^{H}$
(right) of the channel transfer functions for 200 channel realizations.\label{fig:Correlation-matrices-a)}}

\par\end{centering}

\end{figure}

It makes also sense to use a broader definition of correlation, which
is not dependent on the frequency bin, but just on the channel realization.
To this purpose, we consider the deterministic correlation coefficient
$\rho^{H}$, defined as 
\begin{equation}
\rho_{i,j}^{H}=\frac{\sum_{\ell=1}^{L}H_{i}\left(\ell\right)H_{j}^{*}\left(\ell\right)}{\sum_{\ell=1}^{L}\left|H_{i}\left(\ell\right)\right|^{2}\sum_{\ell=1}^{L}\left|H_{j}\left(\ell\right)\right|^{2}},\label{eq:det_corr}
\end{equation}
where $L$ is the total number of frequency bins considered. The results
are plotted in Fig. \ref{fig:Correlation-matrices-a)}, where the
left picture shows the results for $|\rho^{H}|$ and the right one
shows $\rho_{abs}^{H}$, which is the correlation as in \eqref{eq:det_corr}
computed with the absolute value of the transfer functions. On the
main diagonal, instead of plotting the autocorrelation of each channel,
which would be one, we plot $\rho^{H}$ between the Alice to Bob and
the Bob to Alice channels. The results show that the power line channels
are rather uncorrelated (left), and that the correlation increases
when the absolute values of the transfer functions are considered
(right). The correlation between the channels of the legitimate parties
is on average higher than that with Eve, but still not significant
(see Tab. \ref{tab:-of-the}). 

\begin{figure}[tb]
\begin{centering}
\includegraphics[width=1\columnwidth]{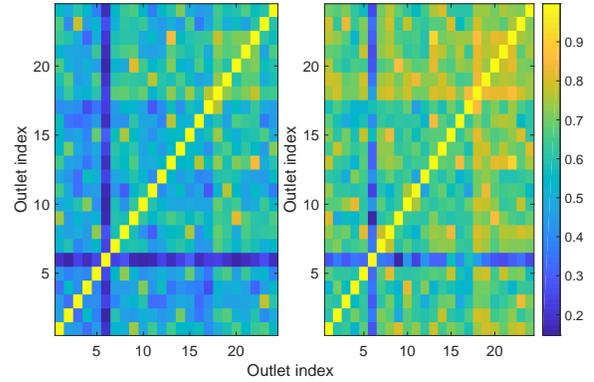}
\par\end{centering}

\caption{Correlation coefficients $|\rho^{Z}|$ (left) and $\rho_{abs}^{Z}$
(right) of the input impedances of 24 outlets in the same household.\label{fig:Impedance-correlation-matrices}}

\end{figure}

In Fig. \ref{fig:Impedance-correlation-matrices}, we plot $\rho^{Z}$
computed as 
\begin{equation}
\rho_{i,j}^{Z}=\frac{\sum_{\ell=1}^{L}Z_{i}\left(\ell\right)Z_{j}^{*}\left(\ell\right)}{\sum_{\ell=1}^{L}\left|Z_{i}\left(\ell\right)\right|^{2}\sum_{\ell=1}^{L}\left|Z_{j}\left(\ell\right)\right|^{2}}.
\end{equation}
The results show that the input impedances are more correlated than
the channel transfer functions, especially when the absolute values
are considered. This might be due to the fact that the input impedance
is, notably at high frequencies, very dependent on the characteristic
impedance of the cable the modem is branched to. If different outlets
are branched to cables with similar characteristic impedances, then
a certain degree of correlation is expected.

\begin{table}[tb]

\caption{$E\left[\rho\right]$ of the channel transfer function in different
cases\label{tab:-of-the}}

\centering{}%
\begin{tabular}{|c|c|c|c|}
\hline 
 & Alice$\leftrightarrow$Bob & Alice$\leftrightarrow$Eve & Ratio\tabularnewline
\hline 
\hline 
CTF \eqref{eq:det_corr} & 0.4452 & 0.1668 & 2.67\tabularnewline
\hline 
CTF absolute values & 0.6298 & 0.4798 & 1.31\tabularnewline
\hline 
Impulse response  & 0.4285 & 0.1147 & 3.74\tabularnewline
\hline 
TDST \eqref{eq:rhoK} & 0.5089 & 0.1808 & 2.81\tabularnewline
\hline 
\end{tabular}
\end{table}

\subsection{Time-domain symmetry technique results\label{sub:Time-domain-symmetry-technique-1}}

\begin{figure}[tb]
\begin{centering}
\subfloat[$\rho_{Alice\leftrightarrow Bob}^{K}$ (dashed) and $\rho_{Alice\leftrightarrow Eve}^{K}$
(solid)\label{fig:8a}]{\centering{}\includegraphics[width=1\columnwidth]{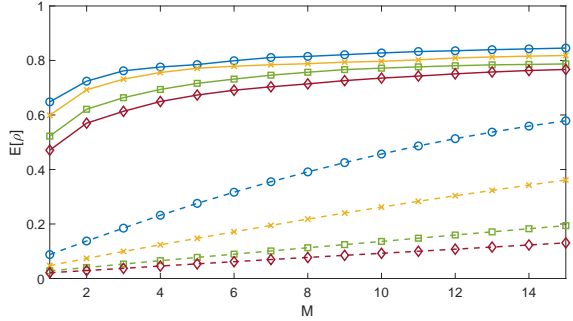}}
\par\end{centering}

\begin{centering}
\subfloat[\label{fig:8b}]{\centering{}\includegraphics[width=1\columnwidth]{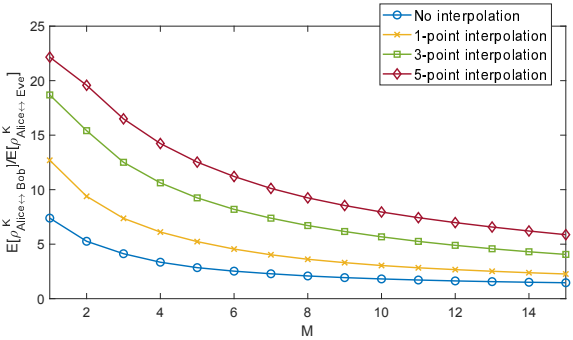}}
\par\end{centering}

\caption{Correlation $R$ of the sequence of peaks computed by Alice, Bob and
Eve (a) and their ratios (b), considering the first $M$ peaks. \label{fig:Correlation-of-the}}
\end{figure}
 As presented in Section \ref{sub:Time-domain-symmetry-technique},
the time-domain channel transfer function is not expected to be more
correlated than the frequency domain one, since $h$ is made by peaks
that have different heights in the two directions. The results when
computing \eqref{eq:det_corr} for the impulse response are similar
to those obtained with the CTF. In fact, the correlation between the
Alice to Bob and Bob to Alice channels is almost the same in the two
cases and only with Eve the correlation is slightly lower in the impulse
response case (see Table \ref{tab:-of-the}).

In order to localize the peaks needed to apply the TDST, we considered
different spectral analysis techniques, both parametric and non-parametric
\cite{stoica2005spectral}. Given the wide bandwidth available, the
best results have been achieved using a non-parametric technique that
consists of interpolating the original estimated time domain trace
and applying the energy based peak detection technique presented in
\cite{6102371}. The interpolation filter is a truncated sinc, which
is equivalent to zero padding in frequency domain. Although the interpolation
does not reveal any more information about the presence of peaks,
it greatly improves the estimation of their location. 

Fig. \ref{fig:Correlation-of-the} shows the average of the correlation
coefficient $\rho^{\bar{h}}$ computed as
\begin{equation}
\rho_{i,j}^{K}=\frac{\sum_{\ell=1}^{L}K_{i}\left(\ell\right)K_{j}^{*}\left(\ell\right)}{\sum_{\ell=1}^{L}\left|K_{i}\left(\ell\right)\right|^{2}\sum_{\ell=1}^{L}\left|K_{j}\left(\ell\right)\right|^{2}},\label{eq:rhoK}
\end{equation}
as a function of the number $M$ of the peaks considered, for different
amounts of interpolation points. As for the length of the key $\bar{h}$,
we set it to $N$=200 elements, which are obtained by segmenting the
impulse response in 200 time blocks, each with duration $\varepsilon=3T_{S}$,
where $T_{S}$ is the sampling period after interpolation. The results
show that when $M$ increases, the correlation between the keys generated
from Alice and Eve increases linearly, while the correlation coefficient
$\rho^{\bar{h}}$ between Alice and Bob almost saturates after the
first steps. This means that considering high values of $M$ for key
generation might reduce the security of the key. On the other hand,
using very low values of $M$ reduces the correlation of the legitimate
parties and also simplifies the work needed by Eve to infer the key
by a series of random guesses. Concerning the interpolation, increasing
the number of interpolation points slightly reduces the correlation
between Alice and Eve, but drastically reduces the one with Eve. This
is particularly clear in Fig. \ref{fig:8b}, which shows that the
ratio between the two correlations is, for example, the same when
using $M=1$ and no interpolation and $M=9$ and 3-point interpolation.
The use of interpolation is therefore encouraged in order to generate
secure keys.

\subsection{Transmission matrix technique results}

\begin{figure}[tb]
\begin{centering}
\includegraphics[width=1\columnwidth]{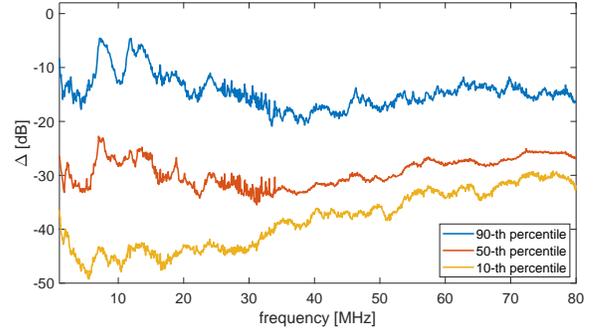}
\par\end{centering}

\caption{Statistical distribution of $\Delta$ in dB scale for different frequencies.\label{fig:Statistical-distribution-of}}
\end{figure}
As explained in Sec. \ref{sub:Transmission-matrix-technique}, the
TMT allows to obtain common CSI between Alice and Bob avoiding Eve
to access it. Since the technique consists of solving a fully determined
system, possible mismatches are only due to the presence of noise.
For example, higher estimation errors are expected when external electromagnetic
interference impinges differently on the two communication ends. It
has been shown in \cite{7476282}, that in the case of PLNs this effect
is limited. Considering for example the key to be generated from $H_{2}$,
Fig. \ref{fig:Statistical-distribution-of} depicts the difference
\begin{equation}
\Delta=\left|\frac{\tilde{H_{2}^{A}}-\tilde{H_{2}^{B}}}{\tilde{H_{2}^{A}}}\right|,
\end{equation}
where $\tilde{H_{2}^{A}}$ is the CTF estimated by Alice and $\tilde{H_{2}^{B}}$
is the one estimated by Bob in the presence of disturbances. The values
of $\Delta$ are on average in the range -25 dB to -35 dB, with one
exception around 10 MHz, where higher values are shown. These higher
values are due to disturbances caused by amateur radio transmissions.

\subsection{Quantization results}

\begin{figure}[tb]
\subfloat[$N$=200\label{fig:10a}]{\begin{centering}
\includegraphics[width=1\columnwidth]{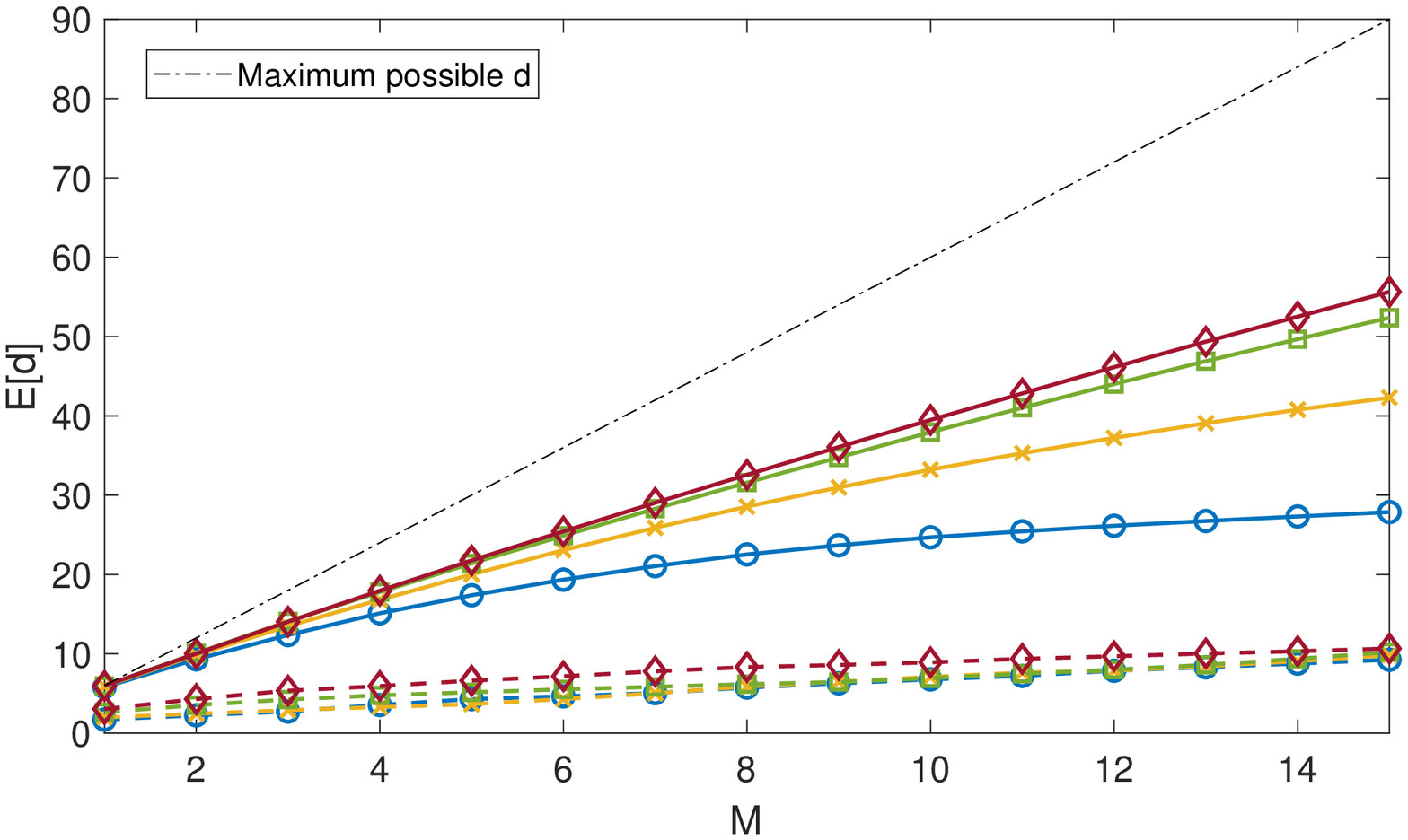}
\par\end{centering}

}

\subfloat[$M$ as high as possible\label{fig:10b}]{\begin{centering}
\includegraphics[width=1\columnwidth]{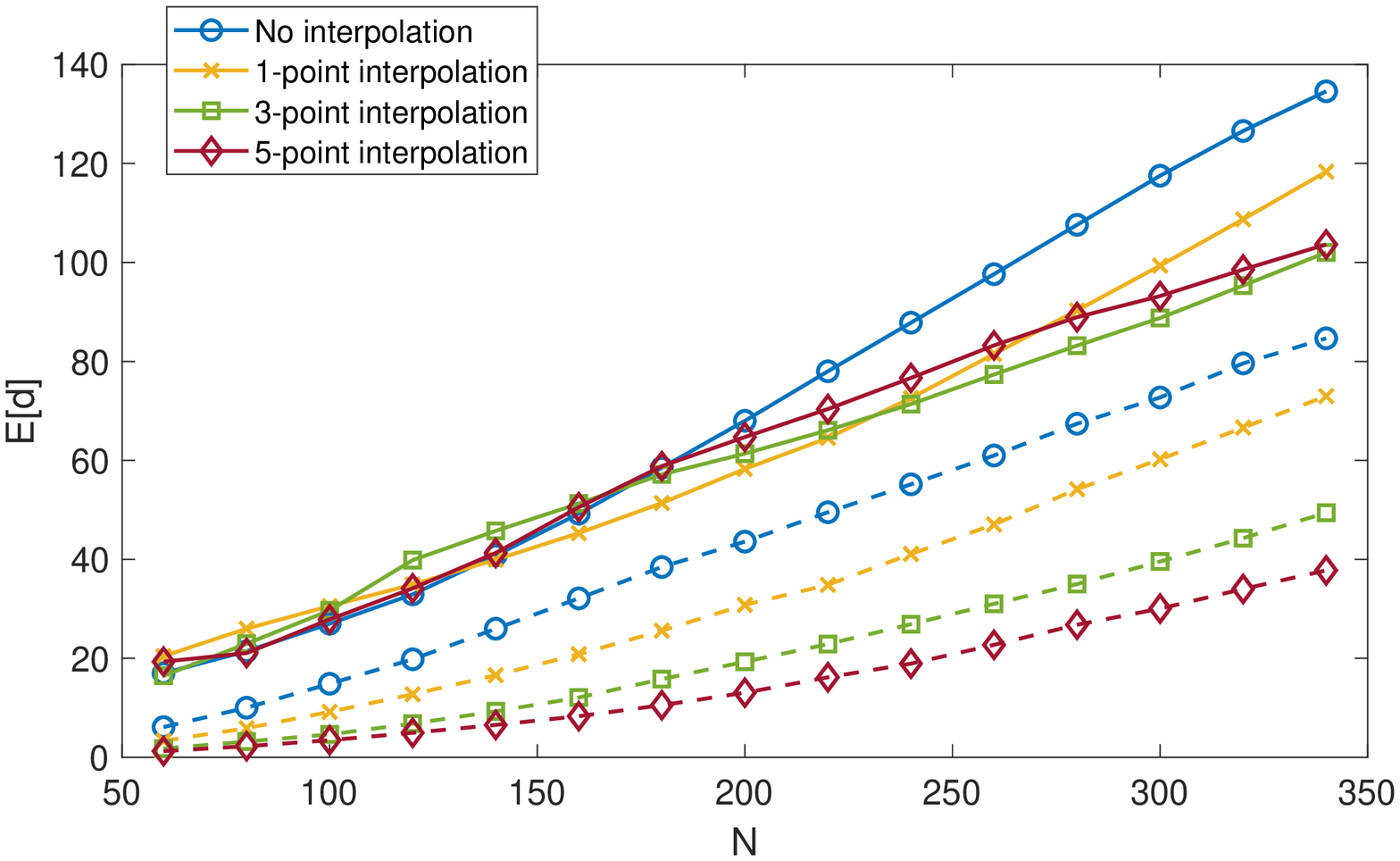}
\par\end{centering}

}\caption{$E\left[d\right]$for different values $M$ of selected peaks (a)
and $N$ of key length (b). Solid and dashed lines refer to the Alice-Eve
and Alice-Bob links, respectively.\label{fig:for-different-values}}

\end{figure}

In order to assess the overall efficiency of the proposed methods,
in this section we analyze the distance $d$ between the keys generated
by the legitimate users and by Eve, using the TDST and the TMT combined
with different quantization methods. We define the distance $d$ between
two keys with equal length $N$ as 

\begin{equation}
d=\frac{\sum_{i=1}^{N}\left|K_{i}^{A}-K_{i}^{B,E}\right|}{\max\left(\max\left(K^{A}\right),\max\left(K^{B,E}\right)\right)},\label{eq:d}
\end{equation}
where $K^{A}$ is the key generated by Alice, $K^{B,E}$ is the key
generated by Bob or Eve. This definition of distance is a normalized
version of the classical Hamming distance \cite{Lint:1982:ICT:601006}.
The two are equal in the binary case, i.e. when $K_{i}^{A},K_{i}^{B,E}\in[0,1]\quad\forall i$.
When the keys are not made of binary symbols, \eqref{eq:d} ensures
that the maximum distance between each symbol is 1. This enables an
easy comparison between different quantization methods over the same
data source.

\begin{figure}[tb]
\subfloat[{$E\left[d_{Alice\leftrightarrow Bob}\right]$ (dashed) and $E\left[d_{Alice\leftrightarrow Eve}\right]$
(solid)\label{fig:11a}}]{\begin{centering}
\includegraphics[width=1\columnwidth]{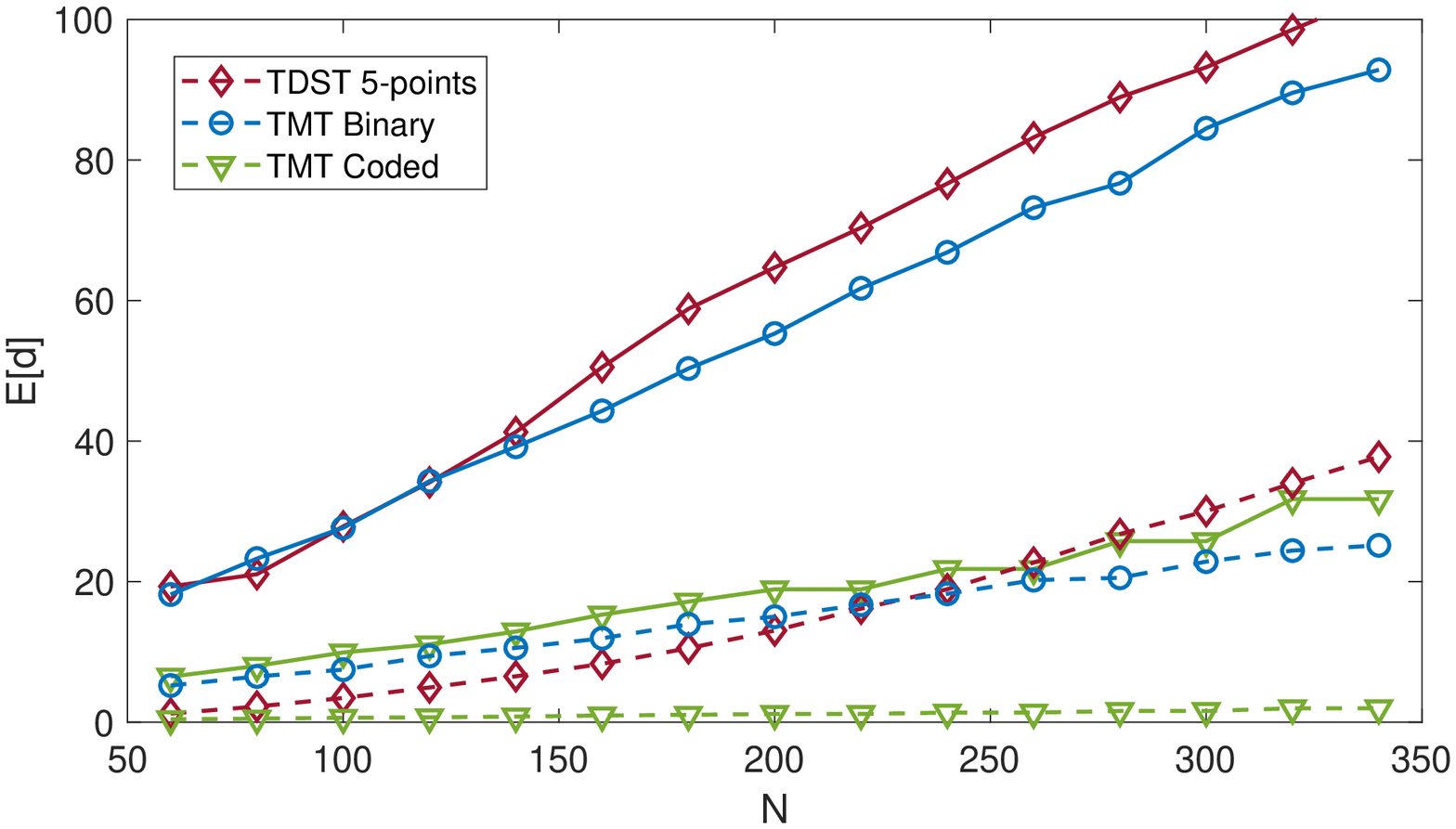}
\par\end{centering}

}

\subfloat[\label{fig:11b}]{\begin{centering}
\includegraphics[width=1\columnwidth]{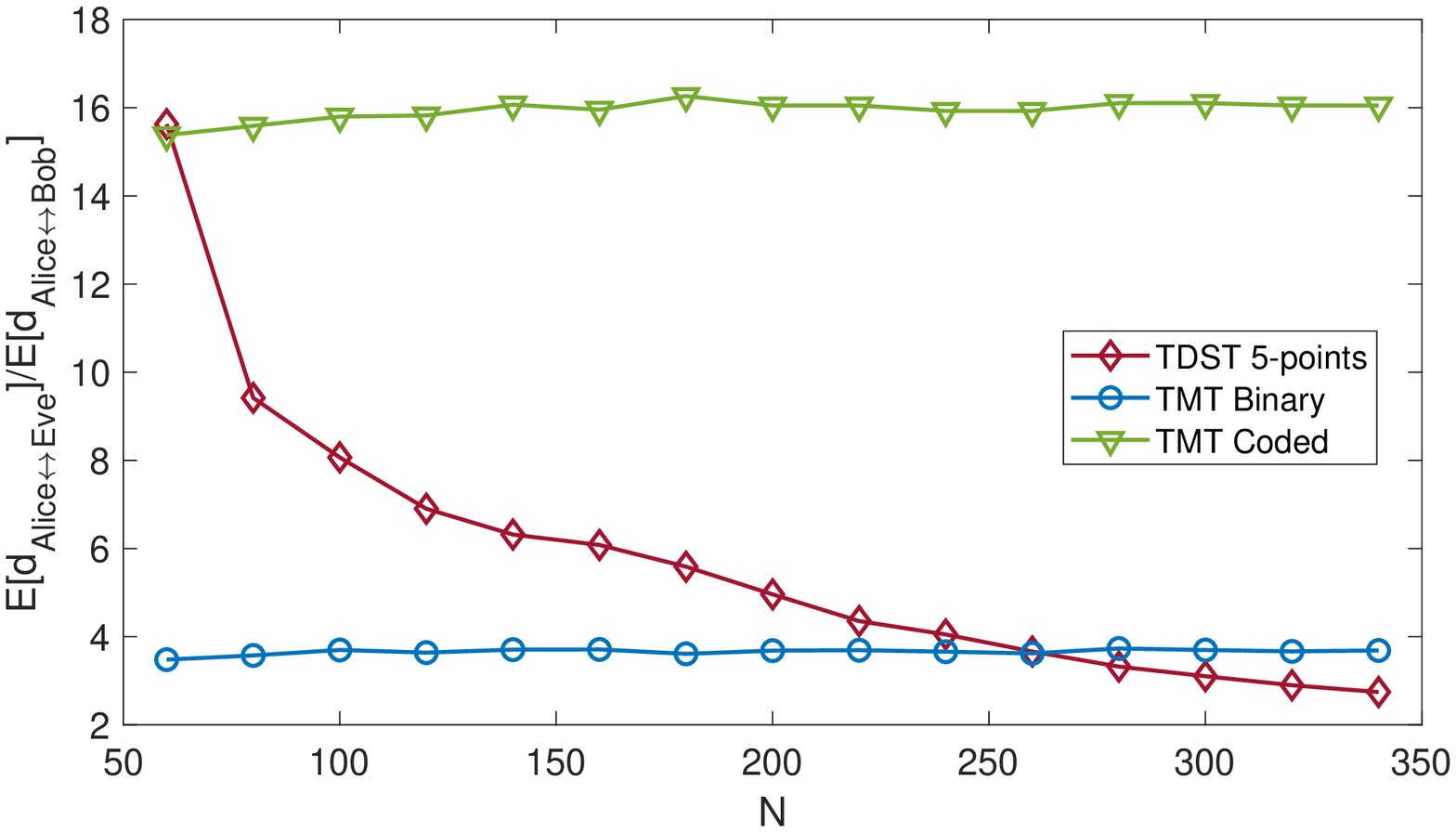}
\par\end{centering}

}\caption{$E\left[d\right]$ of the keys computed by Alice, Bob and Eve (a)
and their ratios (b), for different code lengths $N$ and quantization
methods. The ADC resolution is fixed to 8 bits. \label{fig:-of-the}}
\end{figure}

Fig. \ref{fig:for-different-values} shows the results regarding the
TDST. We notice in Fig. \ref{fig:10a} that as the number $M$ of
bins set to 1 increases, the average $d_{Alice\leftrightarrow Bob}$
only slightly increases and it is almost independent from the interpolation
used. On the other side, the average $d_{Alice\leftrightarrow Bob}$
rapidly detaches from the maximum possible $d$, especially with low
values of interpolation. These results confirm what had already been
deducted from the correlation analysis in Section \ref{sub:Time-domain-symmetry-technique-1}.
Regarding the results in Fig. \ref{fig:10b}, we considered all the
peaks present in $N$ blocks. We notice that in this case $d_{Alice\leftrightarrow Bob}$
is rather influenced by the interpolation factor. This is due to the
fact that, while the first few peaks in the time domain CTF are well
separated and sharp, the density and the smoothness of the other peaks
increases, due to the multipath and the cable attenuation. Therefore,
with increasing $N$ there are much more and less detectable peaks,
which in turn increases $d_{Alice\leftrightarrow Bob}$.

\begin{figure}[tb]
\begin{centering}
\includegraphics[width=1\columnwidth]{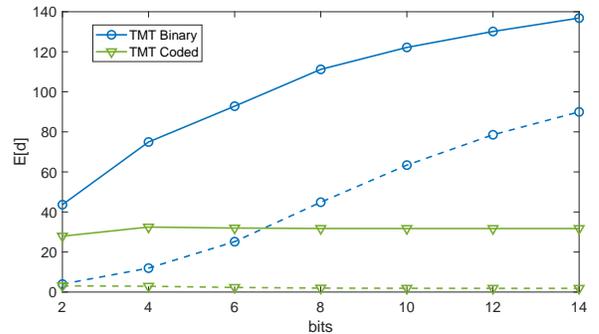}
\par\end{centering}

\caption{$E\left[d\right]$ of the keys computed by Alice, Bob and Eve for
different quantization bits and arranging methods, with $N=340$.
Solid and dashed lines refer to the Alice-Eve and Alice-Bob links,
respectively.\label{fig:-of-the-1}}

\end{figure}

Fig. \ref{fig:-of-the} and \ref{fig:-of-the-1} show the results
regarding the TMT. Fig. \ref{fig:11a} shows that, considering the
same code length $N$, the TMT with binary symbols quantized with
8 bits has a similar performance to the TDST with 5-point interpolation.
On the other hand, the best results in terms of $d_{Alice\leftrightarrow Bob}$
are achieved by the TMT coded method, although also $d_{Alice\leftrightarrow Eve}$
is rather low. However, the ratio between the average $d_{Alice\leftrightarrow Eve}$
and the average $d_{Alice\leftrightarrow Bob}$ is maximized with
this technique (see Fig. \ref{fig:11b}), which therefore ensures
the best security of the key among the proposed methods. 

We finally consider the effect of the number of bits used for quantization
on $d$. As depicted in Fig. \ref{fig:-of-the-1}, while $E[d]$ increases
with the number of bits in the TMT binary case, it is almost independent
from it in the other case. In fact, since the CFT estimated by Alice
and Bob are rather close to each other (cfr Fig. \ref{fig:Statistical-distribution-of}),
the same holds also for the quantized values. When the number of bits
increases, $d$ for each symbol decreases, but at the same time the
number of symbols with non-null $d$ increases, these two effects
compensating each other. In the TMT binary case, on the other hand,
$d$ for each symbol cannot slowly decrease towards 0, since the alphabet
is binary, while the number of symbols with non-null $d$ increases
with $N$.

In conclusion, we found that the best results in term of average $d$
are achieved when considering a limited number of peaks and a high
interpolation factor in the case of the TDST. Regarding the TMT, the
length of the key or the number of quantization levels does not play
a fundamental role, but rather the method to arrange the data. Among
those proposed, the TMT coded yields the best results.

\section{Conclusions\label{sec:Conclusions}}

In this paper, we presented different ways of enhancing physical layer
security in power line networks exploiting the channel properties.
On one side, the power line channel is symmetric when either full
duplex, transresitance or transconductance communication architectures
are used. In this case, the existing methods for physical layer security
in symmetric networks can be applied. On the other side, when the
classical half duplex architecture is used, the power line channel
is not symmetric, but just reciprocal. We showed some fundamental
properties of reciprocal channels that enable the generation of secret
keys with minimal exchange of information between the two legitimate
users. In particular, the wide-sense symmetry of reciprocal channels
has been used to propose a CSI based key generation method that relies
on peak analysis and generates highly correlated information at the
two communication ends with no exchange of key information. Another
CSI based key generation method has been proposed, which relies on
the estimation of the transmission matrix of the link at the two ends
with minimal exchange of information about it through the broadcast
channel. 

We also presented an analysis of the spatial correlation in power
line networks based on a measurement dataset. The results showed that
the power line channels have low spatial correlation, which is even
lower when complex valued CSI is considered. 

We finally generated secret keys by quantizing with different methods
the gathered CSI and assessed their reliability by computing a specifically
formulated distance between the different keys. The results showed
that the distance between the keys generated by Alice and Bob is on
average much lower than the distance between the keys generated by
Alice and Eve. This guarantees a good level of security of the generated
keys.

This paper opens a path for new research efforts in physical layer
security for reciprocal networks. Further developments might include
key agreement protocols, the incorporation of other common information
at the two communication ends and the combination of the proposed
techniques with classical cryptographic methods.

\bibliographystyle{IEEEtran}
\bibliography{femtocell_biblio}

\appendices{}

\section{Wide-sense symmetry of Topology Invariant Channels\label{sec:Wide-sense-symmetry-of}}

According to the Fourier signal theory, every discrete signal $X$
in frequency domain can be written in the form
\begin{equation}
X_{i}=\sum_{k=-\infty}^{+\infty}\Lambda_{k}e^{j\omega_{i}t_{k}},\label{eq:Xi_base}
\end{equation}
where $i$ is the frequency bin index, $j$ is the imaginary unit,
$\omega$ is the radiant frequency, $t$ is the time and $\Lambda$
is a constant. In many application cases, like for the PLC case \cite{SGSI},
the signal can be represented by the sum of $K$ dominant exponentials
as 
\begin{equation}
X_{i}=\sum_{k=1}^{K}\Lambda_{i,k}e^{j\omega_{i}t_{k}},\label{eq:Xi}
\end{equation}
where the minor effects due to the other exponentials are accounted
for in $\Lambda_{i,k}$. In our case, $X$ stands for either $A$,
$B$, $C$ or $D$ in \eqref{eq:ABCD}.  The form of \eqref{eq:Xi}
corresponds to a series of smoothed deltas in time domain, i.e. the
inverse Fourier transform $x_{i}$ of $X_{i}$ is 
\begin{equation}
x_{i}=\sum_{k=1}^{K}\lambda_{i,k}*\delta\left(t-t_{k}\right),\label{eq:xi}
\end{equation}
where $\lambda$ is the inverse Fourier transform of $\Lambda$ and
$*$ is the convolution operator. The delays in \eqref{eq:xi} represent
the time of arrival of each path the signal traveled through. The
values of each $\lambda_{i,k}$ and $t_{k}$ can be derived from the
eigenstructure of the autocorrelation matrix $\mathbf{R_{X}}$ of
$\mathbf{X}=\left[X_{1}\dots X_{N}\right],\quad N>K$, where $N$
is the total number of considered frequency bins \cite{stoica2005spectral}.
In particular, the values of $t_{k}\quad\forall k=1\dots K$, i.e.
the position of the smoothed peaks, are directly derived from the
eigenvectors of $\mathbf{R_{X}}$.

We now consider the CTF of PLC channels. Relying on \eqref{eq:H1},
\eqref{eq:H2} and \eqref{eq:Xi} we can write 
\begin{equation}
H_{1,i}=\frac{1}{\sum_{k=1}^{K}\alpha_{1,i}\Lambda_{i,k}e^{j\omega_{i}t_{k}}}\label{eq:H1i}
\end{equation}
and 
\begin{equation}
H_{2,i}=\frac{1}{\sum_{k=1}^{K}\alpha_{2,i}\Lambda_{i,k}e^{j\omega_{i}t_{k}}},\label{eq:H2i}
\end{equation}
respectively, where $\alpha$ is a multiplicative constant representing
$Z_{T}$, $Z_{L}$ or their product. We remark that, although the
$\alpha$ coefficients are different in \eqref{eq:H1i} and \eqref{eq:H2i},
the exponentials are exactly the same. However, in general $\alpha_{1}$
and $\alpha_{2}$ are different functions of frequency, which in time
domain results in a different peak pattern for $h_{1}$ and $h_{2}$.
Nonetheless, if $Z_{T}$ and $Z_{L}$ are known, $\alpha_{1}$ and
$\alpha_{2}$ can be deconvolved from the CFT. Otherwise, we remark
that in practical cases $Z_{T}$ and $Z_{L}$ are set to constant
values over the band of interest (see Section \ref{sec:Properties-of-the}).
Since the two are proportional to each other, then also $\alpha_{1}$
is proportional to $\alpha_{2}$. This means that $\alpha_{1}$ and
$\alpha_{2}$ have the same propagation modes, and therefore this
holds also for \eqref{eq:H1i} and \eqref{eq:H2i}. 

In conclusion, if $\alpha_{1,i}$ and $\alpha_{2,i}$ are deconvolved
from \eqref{eq:H1i} and \eqref{eq:H2i} or if they are proportional
to each other, $1/H_{1,i}$ and $1/H_{2,i}$ have the same propagation
modes and, therefore, the same sequence of delta functions in time
domain. We define this property as wide-sense time-domain symmetry
of reciprocal systems. According to the Fourier theory, also $H_{1}$
and $H_{2}$ can be written in the form of \eqref{eq:Xi} and, since
their inverse have the same propagation modes, this also holds true
for $H_{1}$ and $H_{2}$.

The wide-sense symmetry can also be explained as follows. Since the
network topology is invariant in PLNs, but also in wireless networks
with the assumption of a time invariant channel, the possible signal
paths between a transmitter and a receiver are fixed. When communication
in the opposite direction is considered, the signal must travel the
same paths, even though in a different direction and in a different
order. These differences cause the signal to receive a different attenuation
in the two directions. But what does not change is the length of each
of those paths. Therefore, the position of the peaks in time domain
is independent from the direction of the communication, but depends
only on the position of the two modems in the network. 

We remark that the wide-sense symmetry does not hold true for $Z_{in1}$
and $Z_{in2}$. In fact, considering \eqref{eq:Zin1} and \eqref{eq:Zin2},
the exponential sequences of $A$, $B$, $C$ and $D$ are combined
through a nonlinear equation that is different for $Z_{in1}$ and
$Z_{in2}$, thus resulting in different propagation modes and peak
sequences in time domain.

\end{document}